\documentclass[final,conference]{IEEEtran}
\usepackage[T1]{fontenc}
\usepackage[latin9]{inputenc}
\usepackage{color}
\usepackage{amsmath}
\usepackage[caption=false,font=footnotesize]{subfig}
\usepackage{bm,amssymb}
\usepackage{amsthm}
\usepackage{pifont}
\theoremstyle{plain}
\newtheorem{thm}{\protect\theoremname}
\theoremstyle{remark}
\newtheorem{rem}[thm]{\protect\remarkname}
\providecommand{\remarkname}{Remark}
\providecommand{\theoremname}{Theorem}

\usepackage{graphicx}

\DeclareMathAlphabet\mathbfcal{OMS}{cmsy}{b}{n} % to describe Tensor

\begin{document}

\title{5G mmWave Downlink Vehicular Positioning}

\author{Henk Wymeersch$^\dagger$,  Nil Garcia$^\dagger$, Hyowon Kim$^*$, Gonzalo Seco-Granados$^\ddagger$,\\
Sunwoo Kim$^*$, Fuxi Wen$^\dagger$, Markus Fr\"{o}hle$^\dagger$\\ \,
$\dagger$ Department of Electrical Engineering, Chalmers University of Technology, Sweden \\
$^*$ Department of Electronic Engineering, Hanyang University, Seoul, Korea \\
$\ddagger$ Department of Telecommunications and Systems Engineering, Universitat Autonoma de Barcelona, Spain 
}
\maketitle
\begin{abstract}
5G new radio (NR) provides new opportunities for accurate positioning from a single reference station: large bandwidth combined with multiple antennas, at both the base station and user sides, allows for unparalleled angle and delay resolution. Nevertheless, positioning quality is affected by multipath and clock biases. We study, in terms of performance bounds and algorithms, the ability to localize a vehicle in the presence of multipath and unknown user clock bias. We find that when a sufficient number of paths is present, a vehicle can still be localized thanks to redundancy in the geometric constraints. Moreover, the 5G NR signals enable a vehicle to build up a map of the environment. 
\end{abstract}

%%%%%%%%%%%%%%%%%%%% N E w   S E C T I ON  %%%%%%%%%%%%%%%%%%%%
\section{Introduction}

Vehicles rely on a variety of sensors to localize themselves and build and maintain a (dynamic) map of the environment \cite{WymSecDesDarTuf:J18}. Most modern cars are equipped with a GPS receiver, which provides absolute location information, and with a radar sensor providing relative location information with respect to detected objects in the environment. GPS operates by estimating pseudoranges with respect to at least four satellites, and solving a system of equations for the user position and clock bias. The main impairments are blocking of GPS signals due to non-line-of-sight (NLOS) and multipath propagation, as signals are reflected on  buildings and other objects \cite{groves2013height}. These impairments lead to positioning errors varying from around one meter in ideal conditions to ten meters or more in GPS-challenged areas, such as urban canyons. In contrast, radar sensors explicitly rely on  multipath propagation: by measuring the transmitted signal reflected from objects, a radar can determine the relative position (bearing and range) with respect to the sensor coordinate frame  \cite{PatTorWaAli17}. %\todo{remove  following? Clock biases are not a challenge in radar, as the transmitter and receiver are co-located and thus share the same clock. However, radar does not directly provide absolute position information.} 

The ability to combine sensing and positioning functionality has recently emerged, mainly in the context of ultra-wide bandwidth communication: with a sufficiently large bandwidth, multipath components become resolvable in the delay domain so that specular paths can be associated with reflectors (or strong scatterers) in the environment \cite{LeitingerJSAC15}. When the locations of reflectors are known, multipath propagation can then help to obtain the position the user. Furthermore, techniques such as simultaneous localization and mapping (SLAM) can be employed to track the user position as well as building up a map of the environment \cite{Durrant-Whyte2006,LeitingerICC2017}. %\todo{ref}.
Conversely, a known user position directly benefits the ability to map the environment. Similar ideas were explored in the context of millimeter-wave communication (mmWave) \cite{GuiMarGue:18}, which will be part of the 5G mobile communication standard. In mmWave, large antenna arrays at the transmitter and receiver provide a high degree of angular resolvability. This means that multipath components can be estimated not only in terms of delay, but also angle. This idea has led to several works on mmWave positioning, which found that a user can process signals from a single base station in order to (i) determine its own position and orientation; (ii) estimate the reflectors in the environment, provided the base station and user were synchronized \cite{Shahmansoori2018,GuerraICCW2015}. % \todo{ref}. 
The synchronization assumption can be relaxed when a two-way protocol is executed between the user and base station, though this leads to additional overheads and challenges in both the uplink and downlink beamforming \cite{Brown2008}. The use of signal strength ranging combined with direction estimation has been also proposed to avoid the need of synchronization \cite{LinLv2018}, but this causes a performance degradation because the large bandwidth of mmWave is not efficiently used for ranging. %\todo{ref}. 

In this paper, we propose to use downlink mmWave signals from a single base station to jointly estimate the vehicle position  and orientation, the environment, and the vehicle's clock bias.  Thereby, the environment is parametrized by the location of virtual anchors (VA) representing specular reflections and scattering of the transmitted signal on objects.
Through a Fisher information analysis, we reveal that multipath propagation is beneficial to estimate the vehicle's clock bias (with respect to the base station), though with some performance penalty compared to a perfectly synchronized scenario. However, in the absence of multipath components, localization of a unsynchronized user using a single base station would be impossible. Moreover, we present a generic downlink positioning system and then analyze specific components of such a system.

%%%%%%%%%%%%%%%%%%%% N E w   S E C T I ON  %%%%%%%%%%%%%%%%%%%%
\section{System Model and Problem Formulation}

\subsection{State Model}
\label{sec:StateModel}
\begin{figure}
\begin{centering}
\includegraphics[width=0.9\columnwidth]{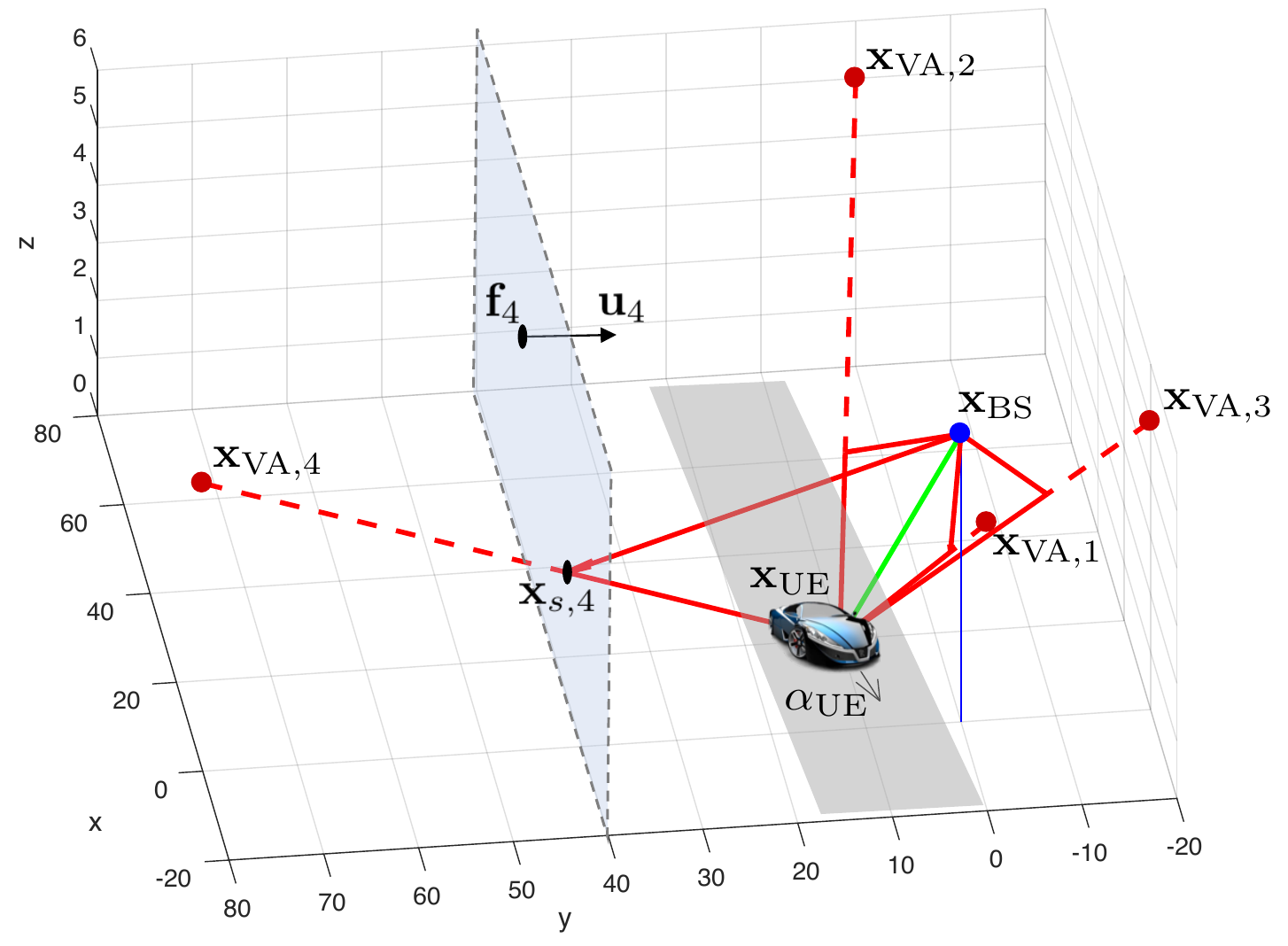}
\par\end{centering}
\caption{Scenario with one base station (blue), one vehicle (heading shown with an arrow), and 4 virtual anchors (each corresponding to a vertical wall). }
\label{fig:scenario}
\end{figure}

We consider a scenario as shown in Fig.~\ref{fig:scenario} with a single static base station (BS), a single mobile user
equipment (UE) mounted on a vehicle, and $M-1$ reflecting surfaces, each
parameterized by a point $\mathbf{f}_{m}$ and a normal vector $\mathbf{u}_{m}$.
The BS is located at $\mathbf{x}_{\mathrm{BS}}=[0,0,z_{\text{BS}}]^{\mathrm{T}}\in\mathbb{R}^{3}$,
so that with each reflecting surface we can associate a virtual anchor 
location \cite{NasKoi17}:
\begin{align}
\mathbf{x}_{\mathrm{VA},m}=\mathbf{P}_{m}\mathbf{x}_{\text{BS}}+\mathbf{t}_{m},
\end{align}
where $\mathbf{P}_{m}=\mathbf{I}_{3}-2\mathbf{u}_{m}\mathbf{u}_{m}^{\mathrm{T}}$
is a Householder matrix and $\mathbf{t}_{m}=2\mathbf{f}_{m}^{\mathrm{T}}\mathbf{u}_{m}\mathbf{u}_{m}$
is a translation vector. Finally, the UE state $\mathbf{s}_{k}=[\mathbf{x}_{\text{UE},k}^{\mathrm{T}}\,\alpha_{\text{UE},k}\,B_{k}]^{\mathrm{T}}$
comprises the vehicle's position, orientation (i.e., the vehicle heading since we consider that the vehicle can only rotate around the vertical axis) and clock
bias, and it is governed by transition function $p(\mathbf{s}_{k} | \mathbf{s}_{k-1})$. 
% * <gonzalo.seco@uab.es> 2018-03-30T17:46:09.326Z:
% 
% I think that we can eliminate this part to shorten the text. In this paper we are processing only one snapshot, so we don't need to include a dynamical model.
% 
% ^ <remero@gmail.com> 2018-04-01T10:01:28.419Z:
% 
% Yes, I agree, maybe next time when we consider tracking, we can use it.
%
% ^.
%where $\mathbf{n}_{k}$ denotes the process noise and $f(\cdot)$ is a transition
%function. 
The epoch duration depends on how frequently the position
is updated. We assume the
UE has a priori information in a factorized form $p(\mathbf{x}_{\mathrm{VA},m})$,
$p(\mathbf{x}_{\text{UE},k})$, $p(\alpha_{\text{UE},k})$ and $p(B_{k})$; and demonstrate in Sec.~\ref{sec:propSolution} how this form can be maintained after updating with measurements. 

\subsection{Measurement Model}

The BS periodically sends a mmWave positioning reference signal (PRS).
At epoch $k$, the received signal at the UE is \cite{Heath2016} %\todo{ref?}
\begin{align}
& \mathbf{y}_{k}(t)=\label{eq:ObservationLow} \\
& \mathbf{W}_{k}^{\mathrm{H}}\sum_{l=0}^{L_k-1}h_{l,k}\mathbf{a}_{\text{UE}}(\bm{\theta}_{l,k})\mathbf{a}_{\text{BS}}^{\mathrm{H}}(\bm{\phi}_{l,k})\mathbf{F}_{k}\mathbf{p}_{k}(t-\tau_{l,k})+\mathbf{W}_k^{\mathrm{H}}\mathbf{n}(t), \nonumber
\end{align}
where  $L_k$ is the number of resolvable propagation paths, \textbf{$\mathbf{F}_{k}$} is a precoder matrix, $\mathbf{p}_{k}(t)$
% * <remero@gmail.com> 2018-04-01T09:59:00.890Z:
% 
% Maybe we should drop subscript k since we are not considering measurements over time.
% 
% ^.
the training signal, \textbf{$\mathbf{W}_{k}$} a combiner matrix, $h_{l}$
is a complex channel gain, $\mathbf{a}_{\text{UE}}$ and $\mathbf{a}_{\text{BS}}$
are the antenna response vectors, and $\tau_{l,k}$, $\bm{\theta}_{l,k}$,
and $\bm{\phi}_{l,k}$ denote time of arrival (TOA), direction of arrival
(DOA), and direction of departure (DOD), respectively, of path $l$ at epoch $k$. Both DOA and DOD have 
azimuth and elevation components. The AWGN at the receiver is denoted $\mathbf{n}(t)$ and has known power spectral density.

%\todo{@Fuxi: make this shorter: 2-3 lines} 
From the observation \eqref{eq:ObservationLow}, several techniques exist to recover the triplet of TOA, DOA, and DOD, such as based on sparse recovery \cite{Shahmansoori2018} or subspace methods \cite{Roemer2014}, which achieve a good balance between estimation accuracy and computational complexity.
Assuming that the $L_k$ propagation paths are
resolvable in the delay and angular domain, a channel estimation % or \todo{and?}
routine provides 
\begin{equation}
\mathbf{z}_{l,k}=[\tau_{l,k},\bm{\theta}_{l,k}^{\mathrm{T}},\bm{\phi}_{l,k}^{\mathrm{T}}]^{\mathrm{T}}+\mathbf{n}_{l,k},\,l \in \{0,1,\ldots,L_k-1\} \label{eq:ObservationHigh}
\end{equation}
where $\mathbf{n}_{l,k}\sim\mathcal{N}(\mathbf{0},\bm{\Sigma}_{l,k})$,
% * <gonzalo.seco@uab.es> 2018-03-30T16:13:19.227Z:
% 
% If we have to reduce the page count, we could start directly with (3) without presenting the received signal model in (2).
% 
% ^.
in which $\bm{\Sigma}_{l,k}$ depends on the channel as well as the
precoding, combining, duration of the training signal, and the receiver. Let the measurement set be $\mathbf{Z}_k=\left\{\mathbf{z}_{l,k}\right\}_{l=0}^{L_k-1}$, where the measurements are unordered as explained below.%, that is, the measurements $\mathbf{z}_{l,k}$ and virtual anchors $\mathbf{x}_{\mathrm{VA},l}$ do not follow the same order. 
% * <gonzalo.seco@uab.es> 2018-03-30T16:08:16.341Z:
% 
% Shouldn't we use another index (e.g. $m$) for the VA's in order to make this absence of ordering more clear?
% 
% ^.

\begin{figure}
\includegraphics[width=1\columnwidth]{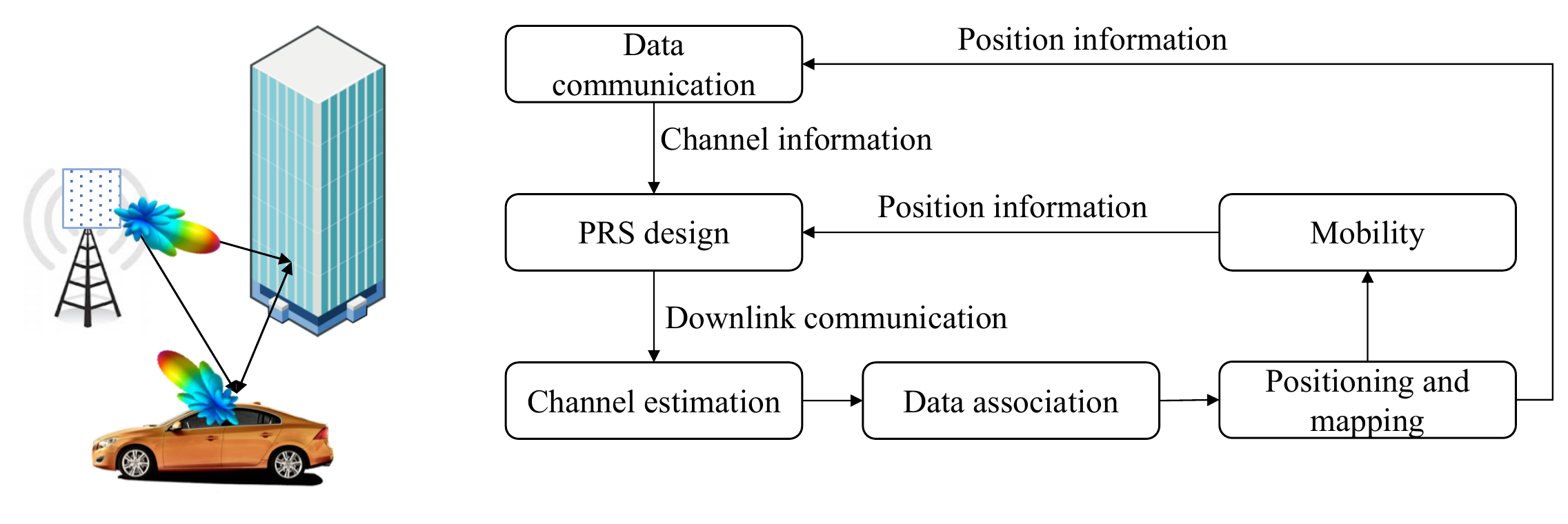}
\caption{The stages of 5G mmWave downlink positioning: the vehicle estimates channel parameters from a dedicated PRS (including precoding and combining), which it associates to prior map information and then uses to refine the vehicle position, heading, and clock bias. 
\label{fig:5G-mmWave-Downlink}}
\end{figure}

\subsection{Problem Formulation }
Our goal is to determine the marginal posterior distributions $p(\mathbf{x}_{\mathrm{VA},m}|\mathbf{Z}_{k})$,
$p(\mathbf{x}_{\text{UE},k}|\mathbf{Z}_{k})$, $p(\alpha_{\text{UE},k}|\mathbf{Z}_{k})$
and $p(B_{k}|\mathbf{Z}_{k})$, given prior distributions on the UE state and possibly some of the VAs. Note that this problem is challenging, due to the unknown clock bias between BS and UE.

\section{Proposed Solution} \label{sec:propSolution}
In this section, we outline the proposed positioning solution, and display the geometric relations between the channel and the location parameters. Since it is not a priori known which measurement in $\mathbf{Z}_k$ corresponds to which VA source, a suboptimal method to deal with this data association problem is presented. Finally, the algorithm to solve the positioning and mapping problem via belief propagation on a factor graph is presented.

\subsection{5G Downlink Positioning}
Our solution approach is shown in Fig.~\ref{fig:5G-mmWave-Downlink}.
First, the mmWave-PRS signals are designed based on prior location information (from the previous epoch, combined with a prediction to the current epoch)
as well as possibly updated information obtained from channel estimation, required during
data transmission. The mmWave-PRS can fill the entire bandwidth in order to localize all users simultaneously\footnote{This is essentially an advantage of downlink positioning. In uplink positioning all users could also also transmit their mmWave-PRS simulatenously and use the entire bandwidth; however, the BS would be forced to separate each signal by applying a spatial filter, making the receiver more complex. Nevertheless, the model \eqref{eq:ObservationHigh} is still valid, with DOA and DOD switching roles. The analysis, data association and positioning can be applied with only minor modification.}  and should be designed for sufficient angular coverage. 
Then, each UE performs channel estimation. Since the
estimates of the $L_k$ paths are not yet tied to the $M$ virtual anchors, a data
association step must follow. Subsequently, the UE performs positioning
and mapping. These estimates can then be provided as inputs for 5G data communication \cite{DiTaranto2014LAC}.
The mobility model (including a model for the clock) is used
to predict the state of the user at the next epoch $k+1$. 

In the following, we will focus on a single epoch and remove the epoch index $k$, with the understanding that the proposed technique should be combined with a Bayesian filter, such as an extended Kalman filter. Moreover,
we will assume that the PRS design and channel estimation are given and
limit our discussion to data association and positioning, starting
from (\ref{eq:ObservationHigh}). 
We are assumed to be provided with (possibly uninformative) prior information about the UE state (in the form
$p(\mathbf{x}_{\mathrm{UE}})$, $p(\alpha_{\mathrm{UE}})$, $p(B)$)
and prior information regarding some VA (in the form $p(\mathbf{x}_{\text{VA},m})$, $m=1,\ldots,M$). The number of paths may be greater or smaller than $M$.  First, we will determine the mapping from the location parameters to the channel parameters. 

\subsection{Relation Between Channel and Location Parameters} \label{sec:relationCHLOC}

Between each virtual anchor $\mathbf{x}_{\text{VA},m}$ and the user position
$\mathbf{x}_{\text{UE}}$, the incidence point of the
specular reflection on the reflecting surface is given by the point where the
straight line between the VA and UE crosses the reflecting surface
(which is itself midway between the BS and the VA), as shown in Fig.~\ref{fig:scenario}. It is
given by 
\begin{equation}
\mathbf{x}_{\text{s},m}=\mathbf{x}_{\text{VA},m}+\frac{(\mathbf{f}_{m}-\mathbf{x}_{\text{VA},m})^{\mathrm{T}}\mathbf{u}_{m}}{(\mathbf{x}_{\text{UE}}-\mathbf{x}_{\text{VA},m})^{\mathrm{T}}\mathbf{u}_{m}}(\mathbf{x}_{\text{UE}}-\mathbf{x}_{\text{VA},m}).\label{eq:IncidencePoint}
\end{equation}
Here, $\mathbf{u}_{m}=(\mathbf{x}_{\text{BS}}-\mathbf{x}_{\text{VA},m})/\Vert(\mathbf{x}_{\text{BS}}-\mathbf{x}_{\text{VA},m})\Vert$
and $\mathbf{f}_{m}=(\mathbf{x}_{\text{BS}}+\mathbf{x}_{\text{VA},m})/2$. Note, this allows to find explicit expressions of $\mathbf{x}_{\text{s},m}$ that only depend
on $\mathbf{x}_{\text{VA},m}$, $\mathbf{x}_{\text{BS}}$, and $\mathbf{x}_{\text{UE}}$ (not shown). Next, we state the relations
between the channel parameters $\tau_{m}$, $\bm{\theta}_{m}=[\theta_{m}^{\text{el}},\theta_{m}^{\text{az}}]^{\mathrm{T}}$,
and $\bm{\phi}_{m}=[\phi_{m}^{\text{el}},\phi_{m}^{\text{az}}]^{\mathrm{T}}$
and the system state. 
\begin{itemize}
\item \emph{Delays:} For the LOS path ($m=0$), $\tau_{0}=\Vert\mathbf{x}_{\text{BS}}-\mathbf{x}_{\mathrm{UE}}\Vert/c+B,$
where $c$ denotes the speed of light. For a NLOS path $m>0$, $\tau_{m}=\Vert\mathbf{x}_{\mathrm{VA},{m}}-\mathbf{x}_{\mathrm{UE}}\Vert/c+B$.\footnote{This is equivalent to $\tau_{m}=\Vert\mathbf{x}_{\mathrm{BS}}-\mathbf{x}_{\text{s},m}\Vert/c+\Vert\mathbf{x}_{\text{s},m}-\mathbf{x}_{\mathrm{UE}}\Vert/c+B$. }
\item \emph{Direction of departure:} For the LOS path, 
\begin{align}
\phi_{0}^{\textrm{az}} & =\arctan\left(\frac{y_{\textrm{UE}}}{x_{\textrm{UE}}}\right)\\
\phi_{0}^{\textrm{el}} & =\arcsin\left(\frac{z_{\textrm{UE}}-z_{\textrm{BS}}}{\|\mathbf{x}_{\mathrm{UE}}-\mathbf{x}_{\mathrm{BS}}\|}\right),
\end{align}
where $\arctan(\cdot)$ is the four-quadrant inverse tangent. Similarly, for the $m$-th NLOS path
\begin{align}
\phi_{m}^{\textrm{az}} & =\arctan\left(\frac{y_{\textrm{s},m}}{x_{\textrm{s},m}}\right)\\ 
\phi_{m}^{\textrm{el}} & =\arcsin\left(\frac{z_{\textrm{s},m}-z_{\textrm{BS}}}{\|\mathbf{x}_{\textrm{s},m}-\mathbf{x}_{\mathrm{BS}}\|}\right).
\end{align}
\item \emph{Direction of arrival:} Here we note that the DOA is measured
in the local frame of reference of the UE, so that the UE orientation must be accounted for. For the LOS path 
\begin{align}
\theta_{0}^{\textrm{az}} & =\pi+\arctan\left(\frac{y_{\textrm{UE}}}{x_{\textrm{UE}}}\right)-\alpha_{\textrm{UE}}\\
\theta_{0}^{\textrm{el}} & =\arcsin\left(\frac{z_{\textrm{BS}}-z_{\textrm{UE}}}{\|\mathbf{x}_{\mathrm{BS}}-\mathbf{x}_{\mathrm{UE}}\|}\right),
\end{align} since the DOA elevation measurement does not depend on the UE orientation, 
while for the $l$-th NLOS path
\begin{align}
\theta_{m}^{\textrm{az}} & =\arctan\left(\frac{y_{\textrm{VA},m}-y_{\textrm{UE}}}{x_{\textrm{VA},m}-x_{\textrm{UE}}}\right)-\alpha_{\textrm{UE}}\\
\theta_{m}^{\textrm{el}} & =\arcsin\left(\frac{z_{\textrm{VA},m}-z_{\textrm{UE}}}{\|\mathbf{p}_{\textrm{VA},m}-\mathbf{p}_{\mathrm{UE}}\|}\right).
\end{align}
\end{itemize}

\subsection{Data Association}

Although the channel estimator provides estimates of
the parameters TOA, DOA, and DOD of each path, it does not reveal which
$\mathbf{z}_{l}$ corresponds to the LOS path and which $\mathbf{z}_{l}$
corresponds to which VA. We consider a simple technique based on the global nearest neighbor assignment \cite{bar1995multitarget}, which provides hard decisions regarding the associations of measurements to VAs.\footnote{A probabilistic/soft data association can also be considered, though this may need modification to the positioning algorithm \cite{meyer2017scalable}.} 
%
%To resolve this assignment,
%we must solve a data association problem. While there are many such
%algorithms, we have consider a simple technique that provides hard
%decisions. 
%
At the current epoch, let $M$ be the number of candidate VAs (including the base station) and $L$ the number of
propagation paths (one per observation vector $\mathbf{z}_{l}$). Each measurement can be explained as coming either from a previously seen VA or as a new VA.\footnote{False alarms due to clutter and missed detections are not considered, for simplicity.}
We create an $L \times (M + L)$ matrix that captures the corresponding likelihoods:
\begin{align}
\mathbf{S}=[\mathbf{S}^\mathrm{D}\,\beta_\mathrm{N}\mathbf{I}_L],
\label{eq:DACostMatrix}
\end{align}
where $\mathbf{I}_L$ is an $L\times L$ identity matrix, %$\mathbf{1}_L$ is a vector of $L$ ones,  $\beta_{\mathrm{F}} \ge 0$ is the false alarm rate, 
$\beta_{\mathrm{N}} \ge 0$ is the new target  rate %, $P_{\mathrm{D}}\in [0,1]$ is the detection probability, 
 and $\mathbf{S}^\mathrm{D}$ is an $L \times M$ matrix, with entries 
\begin{align}
[\mathbf{S}^\mathrm{D}]_{l,m}=p(\mathbf{z}_{l}|{\text{VA}_{m}}),
\end{align}
in which %$P_{\mathrm{D}}\in [0,1]$ is the detection probability and 
% $S=\left[S^\mathrm{D},S^\mathrm{F},S^\mathrm{N}\right]$ with
%  $S^\mathrm{D}_{l,l'}=\frac{P_\mathrm{D}}{1-P_\mathrm{D}}p(\mathbf{z}_{l}|{\text{VA}_{l'}})$ for $l'=1,\ldots,L_\mathrm{VA}$ and $\forall l$, $S^\mathrm{F}=\mathrm{diag}(P_\mathrm{F}\mathbf{1}_L)$, and $S^\mathrm{N}=\mathrm{diag}(P_\mathrm{N} \mathbf{1}_L)$,
%\begin{align}
%& S_{l,l'}= \begin{cases}
%\frac{P_{\mathrm{D}} }{1-P_{\mathrm{D}}}p(\mathbf{z}_{l}|{\text{VA}~l'}) & l' \le %L_{\text{VA}}  \\
%\beta_{\mathrm{F}} & L_{\text{VA}} < l' \le L_{\text{VA}} +L\\
%\beta_{\mathrm{N}}& l' > L_{\text{VA}} +L, 
%\end{cases}
%\end{align}
%where $P_{\mathrm{D}}$ is the detection probability, $\beta_{\mathrm{F}}$ is the false alarm rate, $\beta_{\mathrm{N}}$ is the new target rate, $\mathbf{1}_L$ is an all one vector of dimension $L$, and  
% * <gonzalo.seco@uab.es> 2018-03-30T17:50:39.317Z:
%
% ^.
\begin{align}
& p(\mathbf{z}_{l}|{\text{VA}_{m}}) \approx \mathbb{E}\left\{ p(\mathbf{z}_{l}|\mathbf{x}_{\text{VA},m},\mathbf{x}_{\mathrm{UE}},\alpha_{\mathrm{UE}},B)\right\} \label{eq:costDA}\\
 & =\frac{1}{\sqrt{2\pi|\bm{\Sigma}_{l}|}}\mathbb{E}\left\{ \exp\left(-\frac{1}{2}(\mathbf{z}_{l}-\bm{\eta}_{m})^{\mathrm{T}}\bm{\Sigma}_{l}^{-1}(\mathbf{z}_{l}-\bm{\eta}_{m})\right)\right\}. \nonumber
\end{align}
Here, $\vert\bm{\Sigma}_{l}\vert$ is the determinant of $\bm{\Sigma}_{l}$, which was defined in \eqref{eq:ObservationHigh}, and the function $\bm{\eta}_{m}(\mathbf{x}_{\text{VA},m},\mathbf{x}_{\mathrm{UE}},\alpha_{\mathrm{UE}},B)$
% * <gonzalo.seco@uab.es> 2018-03-30T18:03:26.134Z:
% 
% I think that matrix \bm{\Sigma}_{l} has not been defined.
% 
% ^.
comprises the TOA, DOA, and DOD computed according to Section \ref{sec:relationCHLOC} from
the location of the $m$-th VA (or base station) and UE state. The expectation in \eqref{eq:costDA} can
% * <gonzalo.seco@uab.es> 2018-03-30T19:43:56.121Z:
%
% ^.
easily be computed through Monte Carlo integration from the priors.
Given the matrix $\mathbf{S}$, we then find an optimal assignment by
solving the following optimization problem \vspace{-2.5mm}
% * <gonzalo.seco@uab.es> 2018-03-30T17:55:11.569Z:
%
% ^.
\begin{subequations}
\begin{align}\label{eq:ILP}
\mathrm{maximize~~~} & \sum_{l=0}^{L-1}\sum_{m=1}^{M + L}x_{l,m}\log S_{l,m}\\
\mathrm{s.t.~~~} & x_{l,m}\in\{0,1\},\;\forall l,m,\\
 & \sum_{m=1}^{M + L}x_{l,m}=1,\;\forall l,\\
 & \sum_{l=0}^{L-1}x_{l,m}\le 1,\; \forall m,
\end{align}
\end{subequations}
%\begin{subequations}
%\begin{align}
%\mathrm{maximize~~~} & \sum_{l,l'}x_{l,l'}\log S_{l,l'}\label{eq:ILP}\\
%\mathrm{s.t.~~~} & x_{l,l'}\in\{0,1\}\\
% & \sum_{l'=0}^{L_{\text{VA}} + 2L}x_{l,l'}=1\\
% & \sum_{l=0}^{L-1}x_{l,l'}\le1,
%\end{align}
%\end{subequations}
which can be solved efficiently with the Kuhn-Munkres algorithm \cite{munkres1957algorithms,bourgeois1971extension}. This approach can determine
the LOS path and find new VAs. %The distinction between new VAs and clutter requires data association across multiple epochs, and is not further developed here. 

\begin{rem}
The data association and the localization may be improved by constraining the \emph{joint} probability density function (PDF) of all VA's and the UE state when computing \eqref{eq:costDA}. For instance, when $(\mathbf{x}_\textrm{UE}-\mathbf{x}_{\textrm{s},m})^{\mathrm{T}} (\mathbf{x}_\textrm{BS}-\mathbf{x}_{\textrm{VA},m})<0$ for any $m$, the UE is on the wrong side of the reflecting surface and the joint PDF would be zero. Since our proposed algorithm (see Section \ref{sec:pos_algh}) only provides the \emph{marginal} PDFs, we can account for this  by removing samples that violate such constraints. The association could further be extended to account for gating, when measurements are unlikely with respect to the a priori distribution.  
\end{rem}
\subsection{Positioning and Mapping Algorithm}\label{sec:pos_algh}

\begin{figure}[t!]
\begin{centering}
\includegraphics[width=0.9\columnwidth]{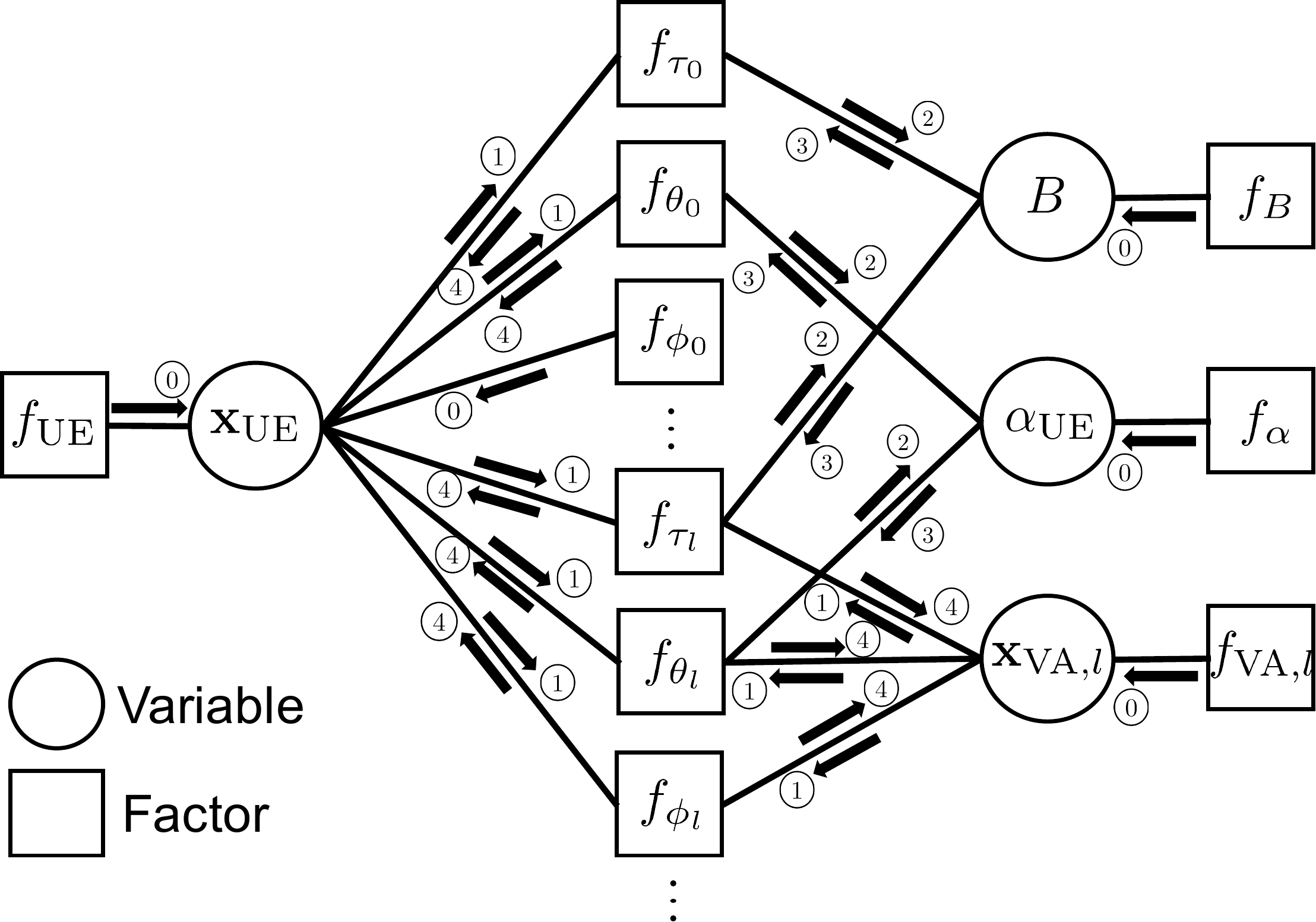}
\par\end{centering}
\caption{Factor graph representation of \eqref{eq:factorization} and the message passing schedule. Here, $f_X$ refers either to the prior of state variable $X$ or the likelihood of the estimate of a channel parameter $X$. }
\label{fig:factor_graph}
\end{figure}
% * <gonzalo.seco@uab.es> 2018-03-30T18:17:10.087Z:
% 
% In the figure with the factor graph:
% - There is one arrow leaving from \alpha_UE without number. I think that it has to be (3).
% - It may be helpful to use the same color for all arrow with (0), another color for the (1)'s, another for the (2)'s ... This could help in visualizing the order of the arrows.
% 
% ^ <khw870511@hanyang.ac.kr> 2018-04-01T11:16:51.221Z:
% 
% Thanks for your comments. The figure of factor graph is fixed.
%
% ^.
Once data association has been performed, we have associated with each measurement $\mathbf{z}_l$ either an existing VA, a new VA (with uniform prior), or a false alarm. Assuming no false alarms, we re-order the VA indices to match the measurement indices. The global posterior distribution can then  be expressed as
\begin{align}
 & p(\mathbf{x}_{\mathrm{UE}},\alpha_{\mathrm{UE}},B,\mathbf{x}_{\mathrm{VA},1},\ldots,\mathbf{x}_{\mathrm{VA},L-1}|\mathbf{Z}) \\
 & =p(\mathbf{x}_{\mathrm{UE}})p(\alpha_{\mathrm{UE}})p(B)\prod_{l=1}^{L-1}p(\mathbf{x}_{\mathrm{VA},l}) \label{eq:factorization} \\ 
 & \times p(\mathbf{z}_{0}|\mathbf{x}_{\mathrm{UE}},\alpha_{\mathrm{UE}},B)\prod_{l=1}^{L-1}p(\mathbf{z}_{l}|\mathbf{x}_{\mathrm{UE}},\alpha_{\mathrm{UE}},\mathbf{x}_{\mathrm{VA},l},B). \nonumber
\end{align}
We make a tacit assumption that factors are removed when needed (e.g., when the data association detects that LOS is not present, the factor with $\mathbf{z}_{0}$ is removed). We aim to compute the marginal posteriors, which can be achieved by executing belief propagation on a factor graph representation of \eqref{eq:factorization}, shown in Fig.~\ref{fig:factor_graph}, where we further approximated $\bm{\Sigma}_l$ from \eqref{eq:ObservationHigh} to have a diagonal structure.\footnote{The proposed technique can be applied, with minor modification, for general $\bm{\Sigma}_l$.} As can be seen, the graph has many cycles, so care needs to be taken when deciding the message passing schedule. In the mmWave regime, we consider accurate measurements of the channel parameters, but possibly uninformative priors on clock bias, UE orientation, as well as UE and VA locations. Our proposed message passing schedule is then as follows. 
% * <gonzalo.seco@uab.es> 2018-03-30T18:04:25.162Z:
% 
% I think that matrix \bm{\Sigma}_{l} has not been defined.
% I imagine that it is the covariance of the channel parameters estimates. 
% But is it diagonal? Didn't we say that there may be correlation between elevation and azimuths estimates (both for DOA and DOD)?
% 
% ^.
\begin{enumerate}
\item[\textcircled{\footnotesize{0}}] The LOS DOD and UE position prior are combined into a message  $\mu_{{\mathbf{x}_{\mathrm{UE}}}}(\mathbf{x}_{\mathrm{UE}}) = \mu_{{f_{\mathbf{x}_\mathrm{UE}}}\rightarrow \mathbf{x}_{\text{UE}}}(\mathbf{x}_{\text{UE}})\mu_{{f_{\mathbf{\phi}_\mathrm{0}}}\rightarrow \mathbf{x}_{\text{UE}}}(\mathbf{x}_{\text{UE}})$. This message captures the knowledge of the UE position based on the LOS DOD (which is always informative, as it is not connected to any other vertex in the graph) and the prior. At the same time all other priors (provided they are informative) send messages to their associated variables (UE bias and orientation, VA position). 
% * <gonzalo.seco@uab.es> 2018-03-30T18:19:46.879Z:
%
% ^.
% \begin{align}
% 	\mu_{{\mathbf{x}_{\mathrm{UE}}}}(\mathbf{x}_{\mathrm{UE}}) = \mu_{{f_{\mathbf{x}_\mathrm{UE}}}\rightarrow \mathbf{x}_{\text{UE}}}(\mathbf{x}_{\text{UE}})\mu_{{f_{\mathbf{\phi}_\mathrm{0}}}\rightarrow \mathbf{x}_{\text{UE}}}(\mathbf{x}_{\text{UE}}). 
% \end{align}
\item[\textcircled{\footnotesize{1}}] The message $\mu_{{\mathbf{x}_{\mathrm{UE}}}}(\mathbf{x}_{\mathrm{UE}})$ is sent to all TOA, DOA, DOD likelihoods for all paths (except DOD for LOS path), and the message $\mu_{{\mathbf{x}_{\mathrm{VA},l}}}(\mathbf{x}_{\mathrm{VA},l})$ is sent to TOA, DOA, DOD likelihoods for $l$-th NLOS path.
\item[\textcircled{\footnotesize{2}}] Each likelihood function, except for DOD, sends a message to bias or orientation. For instance, the message from the TOA likelihood $f_{\tau_l}(\mathbf{x}_{\mathrm{UE}},B,\mathbf{x}_{\mathrm{VA},l})$ to $B$ is given by 
\begin{align}
	 \mu_{f_{\tau_l}\rightarrow B}(B) & = \int \mu_{{\mathbf{x}_{\mathrm{UE}}}}(\mathbf{x}_{\mathrm{UE}})\mu_{\mathbf{x}_{\mathrm{VA},l}\rightarrow f_{\tau_l}}(\mathbf{x}_{\mathrm{VA},l}) \nonumber \\
    & \times f_{\tau_l}(\mathbf{x}_{\mathrm{UE}},B,\mathbf{x}_{\mathrm{VA},l})\mathrm{d}{\sim\{B\}},
\end{align}
where $\mathrm{d}{\sim\{B\}}$ denotes integration over all variables except $B$.
\item[\textcircled{\footnotesize{3}}] Now  the bias and orientation have been updated, they can send a message back to the likelihood functions. For instance, the message from $B$ to $f_{\tau_l}(\mathbf{x}_{\mathrm{UE}},B,\mathbf{x}_{\mathrm{VA},l})$ is given by
\begin{align}
	 \mu_{B \rightarrow f_{\tau_l}}(B) & = 
     p(B) \prod_{l'\neq l}\mu_{f_{\tau_{l'}}\rightarrow B}(B). 
\end{align}
\item[\textcircled{\footnotesize{4}}] All likelihoods, except for LOS DOD, send messages to the UE position variable, and so do NLOS likelihoods  to VA position variable. For instance, the message from $f_{\tau_l}(\mathbf{x}_{\mathrm{UE}},B,\mathbf{x}_{\mathrm{VA},l})$ to $\mathbf{x}_{\mathrm{UE}}$ is given by
\begin{align}
& \mu_{f_{\tau_l}\rightarrow \mathbf{x}_{\mathrm{UE}}}(\mathbf{x}_{\mathrm{UE}}) = \int \mu_{B\rightarrow f_{\tau_l}}(B)  \\
&\times \mu_{\mathbf{x}_{\mathrm{VA},l}\rightarrow f_{\tau_l}}(\mathbf{x}_{\mathrm{VA},l}) 
			f_{\tau_l}(\mathbf{x}_{\mathrm{UE}},B,\mathbf{x}_{\mathrm{VA},l})\mathrm{d}{\sim\{\mathbf{x}_{\mathrm{UE}}\}}. \nonumber
\end{align} Now, the outgoing messages from the UE position and the VA position towards all the likelihood functions can be computed, so we can go back to step \textcircled{\footnotesize{1}}. 
\end{enumerate}
After a sufficient number of iterations between steps \textcircled{\footnotesize{1}}--\textcircled{\footnotesize{4}}, the algorithm is stopped and approximate marginal posteriors are found by multiplication of all incoming messages to the associated variables. For instance, for the UE position, 
\begin{align}
	p(\mathbf{x}_{\mathrm{UE}}|\mathbf{Z})  & \propto p(\mathbf{x}_{\mathrm{UE}}) \prod_{l=0}^{L-1}\mu_{f_{\tau_l}\rightarrow \mathbf{x}_{\mathrm{UE}}}(\mathbf{x}_{\mathrm{UE}})\nonumber\\
			& \times \mu_{f_{\theta_l}\rightarrow \mathbf{x}_{\mathrm{UE}}}(\mathbf{x}_{\mathrm{UE}}) \mu_{f_{\phi_l}\rightarrow \mathbf{x}_{\mathrm{UE}}}(\mathbf{x}_{\mathrm{UE}}),  
\end{align}
and similarly for all the other variables.

\section{Fundamental Performance Analysis}
To gain further understanding in the problem in terms of identifiability and achievable performance, we complement the algorithm description with a Fisher information analysis. 

\subsection{Non-Bayesian Case}
Consider $\bm{\zeta} \in \mathbb{R}^{2+3L}$ to comprise all the location parameters (UE position and orientation and clock bias, VA positions), while $\bm{\eta} \in \mathbb{R}^{5L}$ comprises all the channel parameters. The Fisher information matrix (FIM) of the channel parameters is block diagonal and given by the inverse of the covariance matrix, i.e.,  $\mathbf{J}(\bm{\eta})=\mathrm{blkdiag}\left([ \bm{\Sigma}_{l}^{-1}]_{l=0}^{L-1}\right)$. 
The FIM of the location parameters is found by \cite{Kay1993}
\begin{align}
\mathbf{J}(\bm{\zeta})=\nabla_{\bm{\zeta}}^{\mathrm{T}}\bm{\eta}(\bm{\zeta}) \mathbf{J}(\bm{\eta}(\bm{\zeta})\nabla_{\bm{\zeta}}\bm{\eta}(\bm{\zeta}),
\end{align}
where the Jacobian $[\nabla_{\bm{\zeta}}(\bm{\eta}(\bm{\zeta}))]_{i,j}=\partial [\bm{\eta}(\bm{\zeta})]_i / \partial [\bm{\zeta}]_j$, with the relation $\bm{\eta}(\bm{\zeta}))$ previously explicitly described in Section \ref{sec:relationCHLOC}. The computation of the Jacobian is straightforward but tedious, so it is omitted here for brevity. In case $\mathbf{J}(\bm{\zeta})$ is singular, this means that not all the location parameters are identifiable. 

\subsection{The Use of Prior Information}
When there is prior information $p(\bm{\zeta})$ available, we can compute a hybrid FIM for a fixed $\bm{\zeta}$: %a standard Bayesian FIM (averaged with respect to  $\bm{\zeta}$) or a hybrid FIM for a fixed $\bm{\zeta}$: 
\begin{align}
\mathbf{J}^{\mathrm{hybrid}}(\bm{\zeta})=\mathbf{J}(\bm{\zeta}) + \mathbf{J}^{\mathrm{prior}}, \label{eq:hybridBound}
\end{align}
in which $\mathbf{J}^{\mathrm{prior}}$ comprises the information due to the prior (which is independent on the value of $\bm{\zeta}$). The FIM in \eqref{eq:hybridBound} should be interpreted as obtained from receiving sets of two sources of information: one from the channel estimates and one from the prior.\footnote{Note that the hybrid FIM characterizes the achievable performance for a deterministic estimation problem, where the unknown parameter $\bm{\zeta}$ is fixed, and the estimator uses those two sources of information. It is not a Bayesian FIM, which represents the average achievable performance for the ensemble of  cases where the parameter values are drawn from the distribution of the prior and the estimator exploits the knowledge of this distribution.}  %The use of a prior leads to the parameters always being identifiable (i.e., $\mathbf{J}^{\mathrm{hybrid}}(\bm{\eta})$ is always invertible).

\section{Results}
\subsection{Simulation Environment}
We consider a scenario with 4 vertical walls, a BS at location $[0,0,5]^{\mathrm{T}}$ m, a UE at location $[20, 10, 0]^{\mathrm{T}}$ m and 0 degrees orientation. Virtual anchors are placed in locations $[-20, 0, 5]^{\mathrm{T}}$ m, $[80, 0, 5]^{\mathrm{T}}$ m, $[0, -20, 5]^{\mathrm{T}}$ m,  and $[0,80, 5]^{\mathrm{T}}$ m, as depicted in Fig.~\ref{fig:scenario}. The measurement covariance matrix is set to diagonal, with 0.1 m standard deviation for TOA estimation, and 0.01 rad standard deviation for angle estimation (both DOA and DOD, azimuth and elevation). In terms of prior information, the UE has location standard deviation of 3.2 meters (only in the horizontal plane, perfect knowledge of the vertical coordinate), while the VA 1, 2, 3 have location standard deviation of 10 meters (also in the horizontal plane), except for the VA 4 at location $[0, 80, 5]^{\mathrm{T}}$ m, for which no prior information is available. 
We set $\beta_\mathrm{N} < \min_{l,m} S^{\mathrm{D}}_{l,m}$, 
%$P_\mathrm{D}=1$, $\beta_\mathrm{F}=0$, and $\beta_\mathrm{N}=0.001$, 
needed for the data association \eqref{eq:DACostMatrix}. 
% * <gonzalo.seco@uab.es> 2018-03-30T18:37:45.434Z:
% 
% In the text I was not clear how P_F and P_N are used.
% 
% ^.

\begin{figure}
\begin{centering}
\includegraphics[width=1\columnwidth]{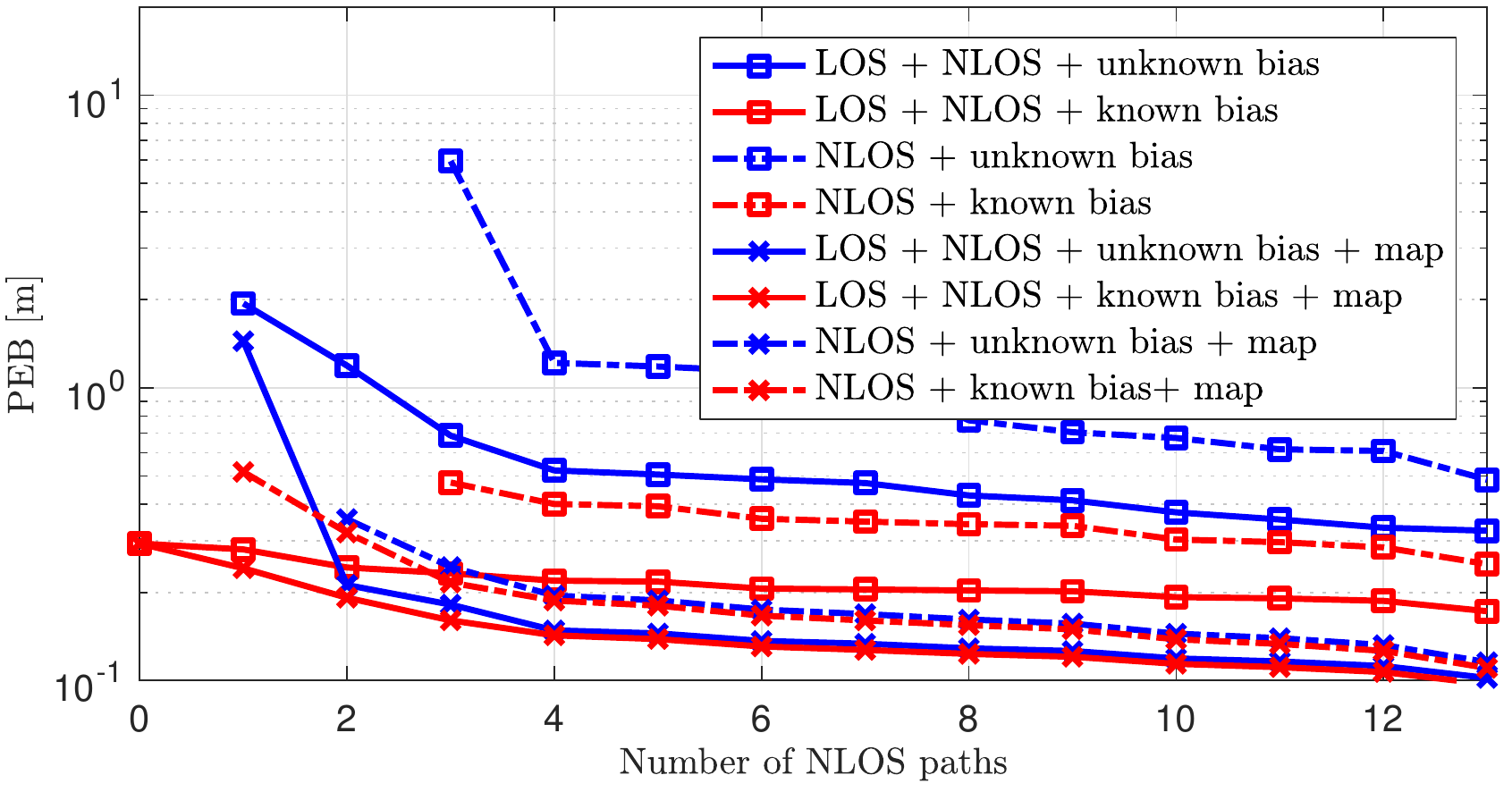}
\par\end{centering}
\caption{PEB as a function of the number of NLOS paths for 8 combinations: with and without a LOS path, with and without knowledge of the map (VA positions), and with and without a known clock bias.}
\label{fig:PEB}
\end{figure}

\begin{figure}
\begin{centering}
\includegraphics[width=1\columnwidth]{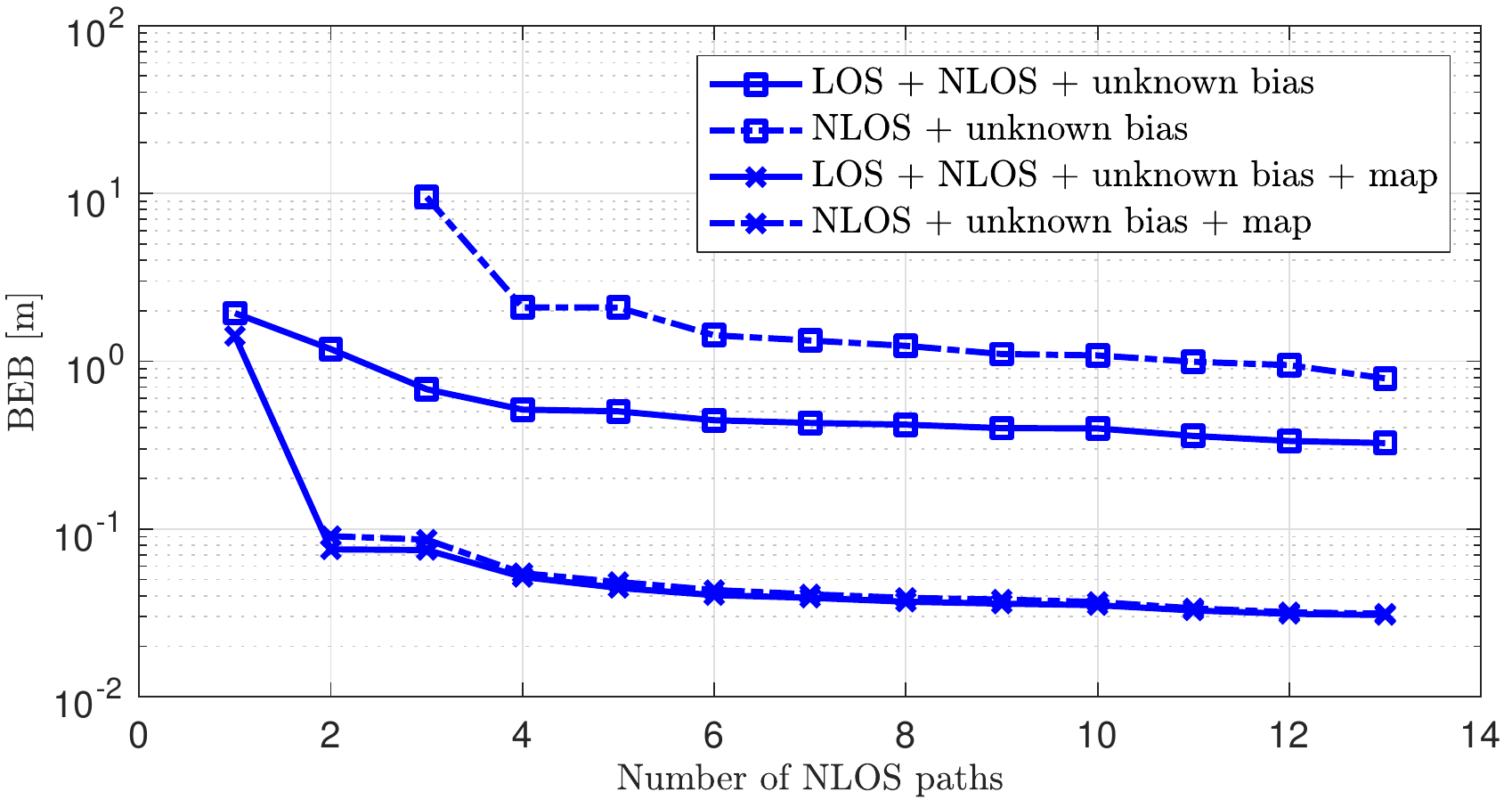}
\par\end{centering}
\caption{BEB as a function of the number of NLOS paths for 4 combinations: with and without a LOS path, with and without knowledge of the map (VA positions).}\label{fig:BEB}
\end{figure}

\begin{figure}
\begin{centering}
\includegraphics[width=1\columnwidth]{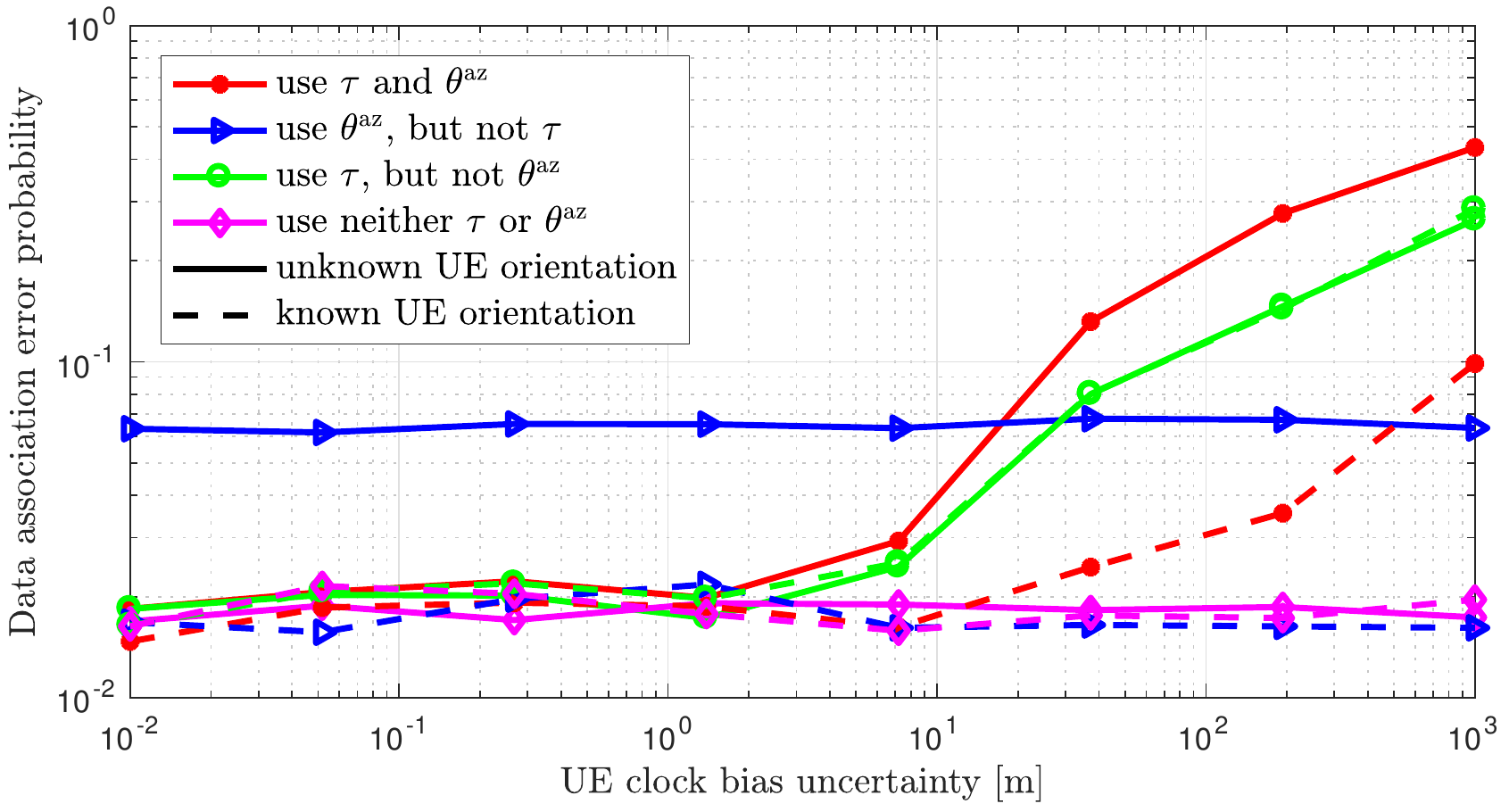}
\par\end{centering}
\caption{Data association error probability as a function of the clock bias uncertainty, when different measurements are used in computing the data association likelihood. Full lines corresponding to no a priori knowledge on the UE orientation, while dashed lines correspond to perfect knowledge of the UE orientation.}\label{fig:DataAssociation}
\end{figure}

\begin{figure*}
\begin{centering}
	\subfloat[\label{Fig:RMSE_UE}]{\includegraphics[width=1\columnwidth]{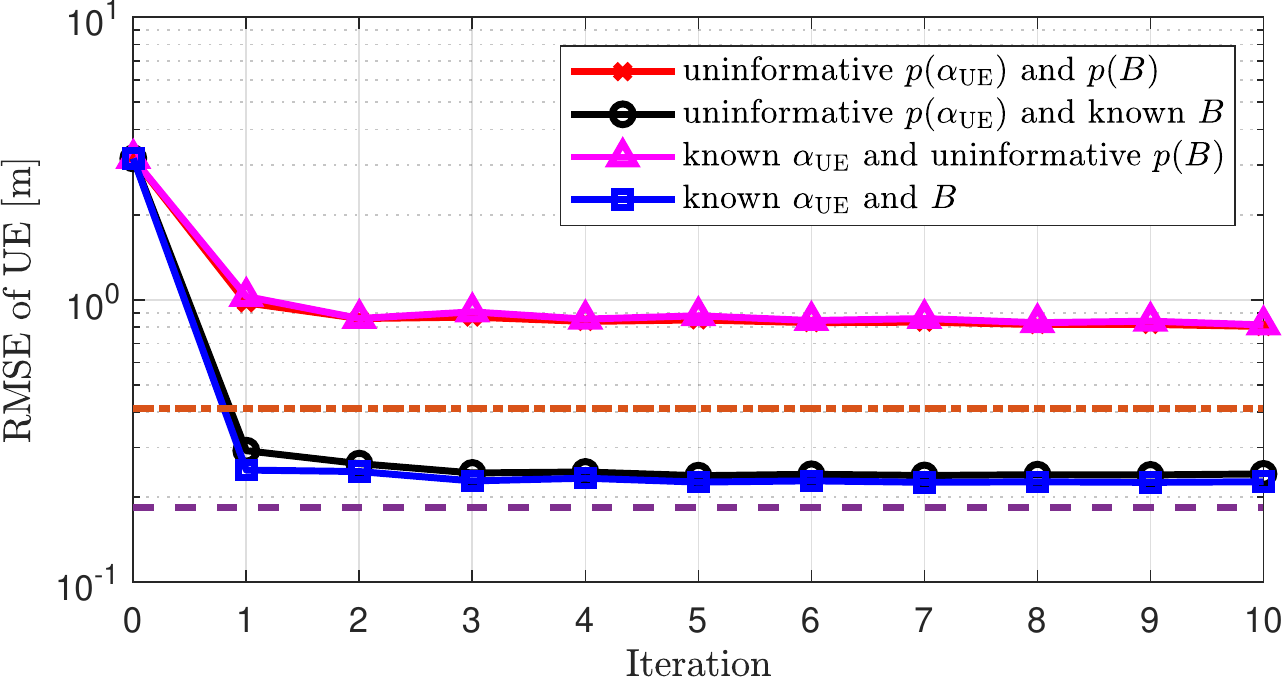}}
	\subfloat[\label{Fig:RMSE_Bias}]{\includegraphics[width=1\columnwidth]{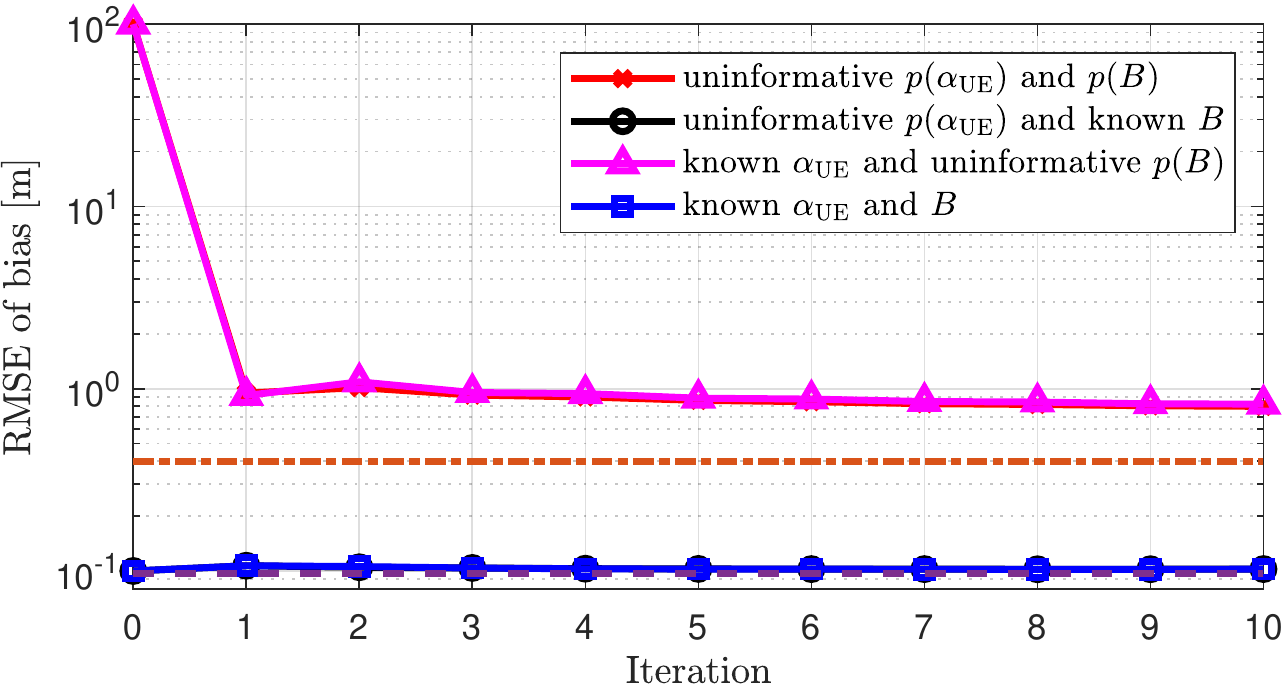}}\\
	\subfloat[\label{Fig:RMSE_Ori}]{\includegraphics[width=1\columnwidth]{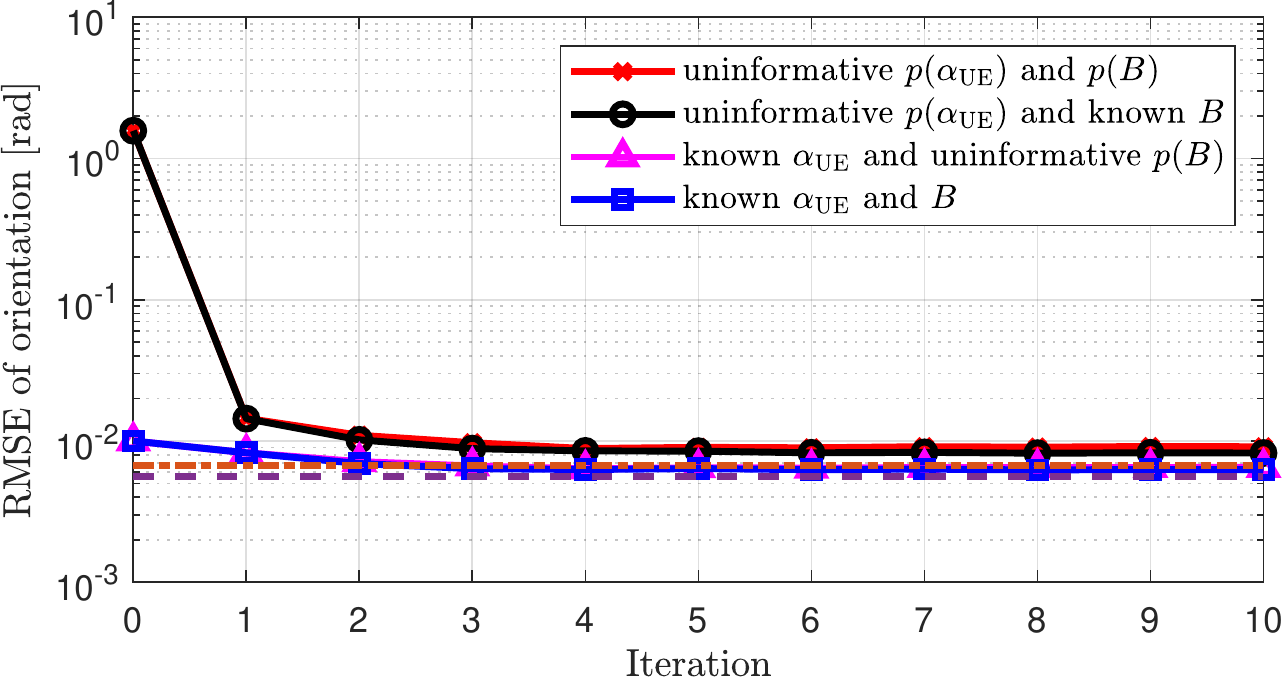}}
	\subfloat[\label{Fig:RMSE_VAs}]{\includegraphics[width=1\columnwidth]{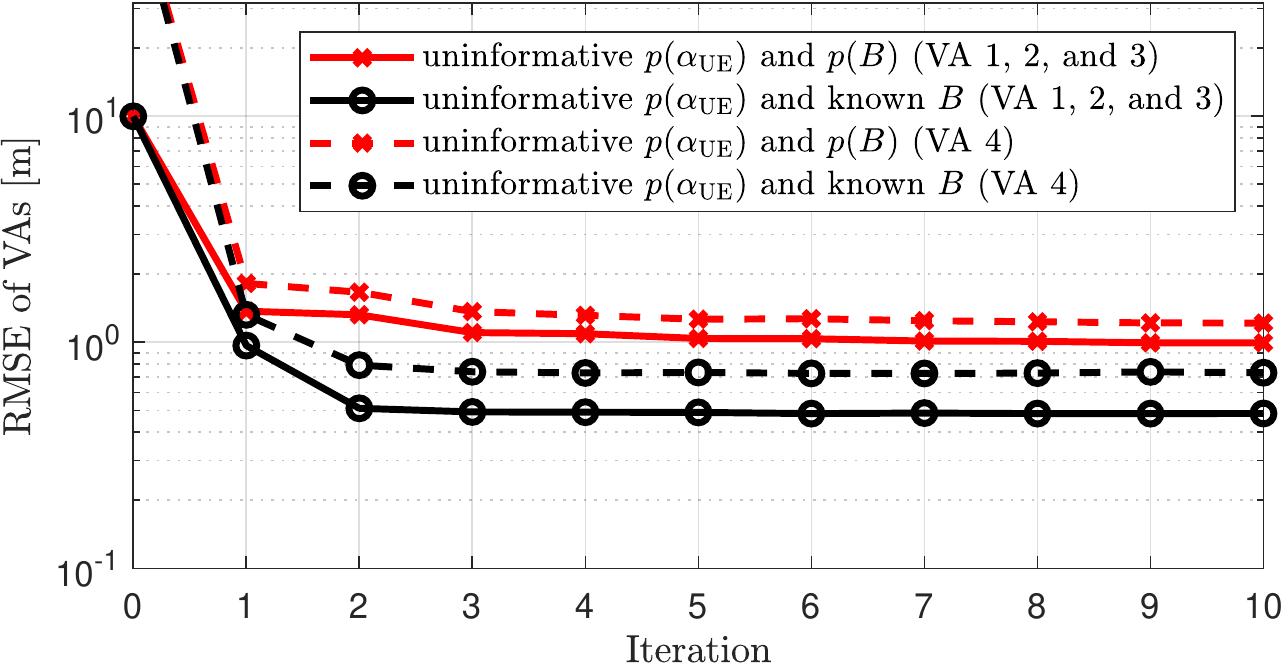}}
	\caption{RMSE as a function of message passing iteration for (a) location of UE, (b)  clock bias, (c) UE orientation, (d) locations of VAs. Horizontal lines are the theoretical performance bounds.}
	\label{Fig:RMSE}
	\par\end{centering}
\end{figure*}
\subsection{FIM Analysis}

We have studied the identifiability of the positioning problem by  considering the scenario above and adding additional virtual anchors (randomly placed). From the inverse of the FIM $\mathbf{J}(\bm{\eta})$, we can derive several quantities of interest, including the position error bound (PEB), the orientation error bound (OEB), the bias error bound (BEB), and the VA error bound (VAEB), which are lower bounds on the achievable accuracy of the position of the UE, the orientation of the UE, the clock bias of the UE, and the VA position, respectively. The PEB is shown as function of the number of NLOS path in Fig.~\ref{fig:PEB}. 
We observe a number of interesting facts. In all cases, having more paths is beneficial. The best performance is achieved when the both LOS and NLOS paths are available, and when the clock bias and VA positions are known (referred to as ``map''), while the worst performance is achieved when only NLOS paths are available, and neither the clock bias nor the VA positions are known. Provided enough paths are available, the system state is always identifiable in spite of the fact that the UE has not a synchronized clock. With an unknown clock bias, one NLOS path is needed when LOS is present. When LOS is not present, at least three NLOS paths are needed, or only two in case map information is available (i.e., the position of the VAs). These results are corroborated by Fig.~\ref{fig:BEB}, which clearly confirms that the clock bias can be estimated even with one-way transmission as long as the scenario provides enough diversity in terms of NLOS paths. In the absence of map information, the PEB for the scenario from Fig.~\ref{fig:scenario} ranges from 2.5 m (NLOS only, unknown bias), over 1 m (LOS and NLOS, unknown bias), to 40 cm (LOS and NLOS, known bias); and the BEB ranges from 2 m (NLOS only, unknown bias) to 0.5 m (LOS and NLOS, unknown bias), while the availability of map information reduces the bounds in one order of magnitude or more.

% \begin{figure}
% \begin{centering}
% \includegraphics[width=1\columnwidth]{OEB-crop.pdf}
% \par\end{centering}
% \caption{OEB}
% \end{figure}

% \begin{figure}
% \begin{centering}
% \includegraphics[width=1\columnwidth]{BEB-crop.pdf}
% \par\end{centering}
% \caption{BEB}
% \end{figure}

\subsection{Data Association}
Once channel parameter estimates are available, the  expectation in \eqref{eq:costDA} is computed using Monte Carlo integration with 1000 samples. We evaluate the data association performance in terms of the data association error probability (i.e., the probability that a measurement is incorrectly assigned) as a function of the a priori uncertainty in the clock bias. The optimization problem \eqref{eq:ILP} is solved using the Kuhn-Munkres algorithm, which has complexity $\mathcal{O}(L^3)$. Uncertainty in the clock bias and UE orientation affects the TOA measurements and azimuth DOA measurements, respectively. We investigate the impact on the data association when these measurements are included or not in the computation of the data association likelihood. Fig.~\ref{fig:DataAssociation} shows the corresponding results. When the clock bias is known well (i.e., less than 1 meter uncertainty) it is beneficial to include the TOA measurements. However, when the clock bias is highly uncertain, the best performance is achieved when both TOA and azimuth DOA measurements are ignored. In conclusion, channel parameters that relate to highly uncertain parameters should be avoided during data association. 
% * <gonzalo.seco@uab.es> 2018-03-30T19:21:21.198Z:
% 
% It is not clear to what n means in $\mathcal{O}(n^3)$.
% 
% ^.

% * <gonzalo.seco@uab.es> 2018-03-30T20:29:59.925Z:
% 
% Maybe we can reduce this figure to two subplots and this will help in reaching the 6 pages.
% 
% ^.
\subsection{Localization Performance}
We study the localization performance after (perfect) data association, using the algorithm described in Section \ref{sec:pos_algh}, in four different cases: the prior for the bias $B$ is informative (standard deviation 0.1 m) or uninformative (standard deviation 100 m); the prior for the orientation $\alpha_\mathrm{UE}$ is informative (standard deviation 0.01 rad) or uninformative (standard deviation $\pi/2$). We use {2000} samples for approximating the posterior distributions and perform 10 iterations of message passing. After 200 Monte Carlo runs, we evaluate the root mean squared error (RMSE) of the UE position, the UE bias, the UE orientation, and the VA location, as a function of the iteration index. The results are shown in Fig.~\ref{Fig:RMSE}, including the hybrid bounds derived from \eqref{eq:hybridBound} (horizontal lines). We observe that in all cases, the algorithm can operate close to the bound and converges after {3 or 4} iterations. It is clearly shown in Fig.~\ref{Fig:RMSE_UE} that knowing $B$ is in general more beneficial than knowing $\alpha_\mathrm{UE}$, as the latter parameter can be accurately estimated from the LOS path, while the former cannot. Fig.~\ref{Fig:RMSE_VAs} demonstrates that having prior information yields better localization accuracy for VA 1--3, compared to VA 4. The final performance under uninformative priors on the clock bias and UE orientation is approximately {80} cm UE error, {80} cm clock bias error ({2.6} ns), {0.01} rad orientation error, and {0.5} m (resp. {0.75} m) VA location error for VAs with informative prior (resp. uninformative prior).

\section{Conclusion}
5G mmWave signals have unique properties for precise positioning of vehicles and can complement existing on-board sensors. We remove the common assumption of synchronization between vehicle and base station by proposing a framework for joint estimation of the vehicle's position and heading, as well as the clock bias. The method relies on the ability to resolve multipath components and estimate their angles and delays, so as to build up a map of the propagation environment. A Fisher information analysis reveals that the LOS and only one NLOS paths are sufficient to estimate all parameters. In case that the LOS is not present, then three NLOS paths are needed. The performance is also evaluated through a belief propagation method, which is able to attain the performance bounds. 

\section{Acknowledgments}
This work was supported, in part, by the EU H2020 projects HIGHTS MG-3.5a-2014-636537 and 5GCAR, and the VINNOVA COPPLAR project, funded under Strategic Vehicle Research and Innovation Grant No. 2015-04849, the Ministry of Science, ICT and Future Planning, Korea, under the ITRC support program (IITP-2018-0-01637) supervised by the Institute for Information \& communications Technology Promotion, Samsung Research Funding \& Incubation Center of Samsung Electronics under Project Number SRFC-IT-1601-09, the Spanish Ministry of Economy, Industry and Competitiveness, under Grant TEC2017-89925-R. We are also thankful to Rico Mendrzik for providing valuable feedback.
\bibliographystyle{IEEEtran}
\bibliography{IEEEabrv,references}

\begin{thebibliography}{10}
\providecommand{\url}[1]{#1}
\csname url@samestyle\endcsname
\providecommand{\newblock}{\relax}
\providecommand{\bibinfo}[2]{#2}
\providecommand{\BIBentrySTDinterwordspacing}{\spaceskip=0pt\relax}
\providecommand{\BIBentryALTinterwordstretchfactor}{4}
\providecommand{\BIBentryALTinterwordspacing}{\spaceskip=\fontdimen2\font plus
\BIBentryALTinterwordstretchfactor\fontdimen3\font minus
  \fontdimen4\font\relax}
\providecommand{\BIBforeignlanguage}[2]{{%
\expandafter\ifx\csname l@#1\endcsname\relax
\typeout{** WARNING: IEEEtran.bst: No hyphenation pattern has been}%
\typeout{** loaded for the language `#1'. Using the pattern for}%
\typeout{** the default language instead.}%
\else
\language=\csname l@#1\endcsname
\fi
#2}}
\providecommand{\BIBdecl}{\relax}
\BIBdecl

\bibitem{WymSecDesDarTuf:J18}
H.~Wymeersch, G.~Seco-Granados, G.~Destino, D.~Dardari, and F.~Tufvesson,
  ``{5G} mm-{W}ave positioning for vehicular networks,'' \emph{Wireless Commun.
  Mag.}, vol.~24, no.~6, pp. 80--86, Dec. 2018.

\bibitem{groves2013height}
P.~D. Groves and Z.~Jiang, ``Height aiding, ${C}/{N}_0$ weighting and
  consistency checking for {GNSS} {NLOS} and multipath mitigation in urban
  areas,'' \emph{J. Navigat.}, vol.~66, no.~5, pp. 653--669, Sep. 2013.

\bibitem{PatTorWaAli17}
S.~M. Patole, M.~Torlak, D.~Wang, and M.~Ali, ``Automotive radars: A review of
  signal processing techniques,'' \emph{IEEE Signal Process. Mag.}, vol.~34,
  no.~2, pp. 22--35, Mar. 2017.

\bibitem{LeitingerJSAC15}
E.~Leitinger, P.~Meissner, C.~Ruedisser, G.~Dumphart, and K.~Witrisal,
  ``Evaluation of position-related information in multipath components for
  indoor positioning,'' \emph{{IEEE} J. Sel. Areas Commun.}, vol.~33, no.~11,
  pp. 2313 -- 2328, Nov. 2015.

\bibitem{Durrant-Whyte2006}
H.~Durrant-Whyte and T.~Bailey, ``Simultaneous localization and mapping: part
  {I},'' \emph{IEEE Robot. Autom. Mag.}, vol.~13, no.~2, pp. 99--110, Jun.
  2006.

\bibitem{LeitingerICC2017}
E.~Leitinger, F.~Meyer, F.~Tufvesson, and K.~Witrisal, ``Factor graph based
  simultaneous localization and mapping using multipath channel information,''
  in \emph{Proc. IEEE ICC-17}, Paris, France, Jun. 2017.

\bibitem{GuiMarGue:18}
F.~Guidi, A.~Mariani, A.~Guerra, D.~Dardari, A.~Clemente, and R.~D'Errico,
  ``Indoor environment-adaptive mapping with beamsteering massive arrays,''
  \emph{IEEE Transactions on Vehicular Technology}, 2018.

\bibitem{Shahmansoori2018}
A.~Shahmansoori, G.~E. Garcia, G.~Destino, G.~Seco-Granados, and H.~Wymeersch,
  ``Position and orientation estimation through millimeter-wave {MIMO} in 5{G}
  systems,'' \emph{IEEE Trans. Wireless Commun.}, vol.~17, no.~3, pp.
  1822--1835, Mar. 2018.

\bibitem{GuerraICCW2015}
A.~Guerra, F.~Guidi, and D.~Dardari, ``Position and orientation error bound for
  wideband massive antenna arrays,'' in \emph{Proc. IEEE ICC Workshop on
  Advances in Network Localization and Navigation (ANLN)}, Jun. 2015, pp.
  853--858.

\bibitem{Brown2008}
D.~R. {Brown III} and H.~V. Poor, ``Time-slotted round-trip carrier
  synchronization for distributed beamforming,'' \emph{IEEE Trans. Signal
  Process.}, vol.~56, no.~11, pp. 5630--5643, Nov. 2008.

\bibitem{LinLv2018}
Z.~Lin, T.~Lv, and P.~T. Mathiopoulos, ``{3-D} indoor positioning for
  millimeter-wave massive {MIMO} systems,'' \emph{IEEE Trans. Commun.},
  vol.~PP, no.~99, pp. 1--1, Jan. 2018.

\bibitem{NasKoi17}
H.~Naseri and V.~Koivunen, ``Cooperative simultaneous localization and mapping
  by exploiting multipath propagation,'' \emph{IEEE Transactions on Signal
  Processing}, vol.~65, no.~1, pp. 200--211, Jan 2017.

\bibitem{Heath2016}
R.~W. Heath, N.~Gonz\'alez-Prelcic, S.~Rangan, W.~Roh, and A.~M. Sayeed, ``An
  overview of signal processing techniques for millimeter wave {MIMO}
  systems,'' \emph{IEEE J. Sel. Topics Signal Process.}, vol.~10, no.~3, pp.
  436--453, Apr. 2016.

\bibitem{Roemer2014}
F.~Roemer, M.~Haardt, and G.~D. Galdo, ``Analytical performance assessment of
  multi-dimensional matrix- and tensor-based {ESPRIT}-type algorithms,''
  \emph{IEEE Trans. Signal Process.}, vol.~62, no.~10, pp. 2611--2625, May
  2014.

\bibitem{DiTaranto2014LAC}
R.~D. Taranto, S.~Muppirisetty, R.~Raulefs, D.~Slock, T.~Svensson, and
  H.~Wymeersch, ``Location-aware communications for {5G} networks: How location
  information can improve scalability, latency, and robustness of {5G},''
  \emph{IEEE Signal Processing Magazine}, vol.~31, no.~6, pp. 102--112, Nov
  2014.

\bibitem{bar1995multitarget}
Y.~Bar-Shalom and X.-R. Li, \emph{Multitarget-multisensor tracking: principles
  and techniques}.\hskip 1em plus 0.5em minus 0.4em\relax YBS, UK, 1995.

\bibitem{meyer2017scalable}
F.~Meyer, P.~Braca, P.~Willett, and F.~Hlawatsch, ``A scalable algorithm for
  tracking an unknown number of targets using multiple sensors,'' \emph{IEEE
  Trans. Signal Process.}, vol.~65, no.~13, pp. 3478--3493, 2017.

\bibitem{munkres1957algorithms}
J.~Munkres, ``Algorithms for the assignment and transportation problems,''
  \emph{J. Soc. Indust. Appl. Math.}, vol.~5, no.~1, pp. 32--38, Mar. 1957.

\bibitem{bourgeois1971extension}
F.~Bourgeois and J.-C. Lassalle, ``An extension of the munkres algorithm for
  the assignment problem to rectangular matrices,'' \emph{Commun. ACM},
  vol.~14, no.~12, pp. 802--804, Dec. 1971.

\bibitem{Kay1993}
S.~Kay, \emph{Fundamentals of{ Statistical Signal Processing: Estimation
  Theory}}.\hskip 1em plus 0.5em minus 0.4em\relax Prentice Hall Signal
  Processing Series, 1993.

\end{thebibliography}

\end{document}